\documentclass[]{JHEP3}
\usepackage{amsmath,cite}

\newcommand{\eqn}[1]{(\ref{#1})}

\title{Chiral Supergravity}
\author{Melanie Becker\thanks{mbecker@physics.tamu.edu}, Paul Bruillard\thanks{pjb2357@physics.tamu.edu}, and Sean Downes\thanks{sddownes@physics.tamu.edu}\\
George P. and Cynthia W. Mitchell Institute for Fundamental Physics\\
Texas A$\&$M University, College Station, TX 77843, USA}

\abstract{
We study the linearized approximation of ${\cal N}=1$ topologically
massive supergravity around $AdS_{3}$. Linearized gravitino fields
are explicitly constructed. For appropriate boundary conditions, the
conserved charges demonstrate chiral behavior, so that chiral
gravity can be consistently extended to Chiral Supergravity.}


\preprint{MIFP-08-13\\ hep-th//yymmnnn}

\begin{document}
\section{Introduction and Motivation}
\label{introduction section}

The immense number of vacua in the String Theory Landscape
is surrounded by an even larger number of vacua contained in the
Swampland \cite{Vafa}. That is, the set of effective theories that appear valid semiclassically, but are inconsistent quantum theories. One might wonder
if, in the gravitational scenery, there could exist a third class
of theories: renormalizable and fully consistent quantum theories of
gravity that stand independent of String Theory. Candidate theories include
${\cal N}=8$ Supergravity in four dimensions (see \cite{Bern} for the latest status) and
Supergravity in three dimensions. In this paper, we are interested
in the latter.
\ \\
\ \\
In three dimensions, gravity is highly constrained, suggesting that some theories might be consistent at the quantum level.
Chiral Gravity \cite{andy1}, Log Gravity \cite{grjo1,Maloney,AyonBeato1,AyonBeato2}, and New Massive Gravity \cite{behoto} may be
examples of such theories, although the second is expected to
have a non-unitary dual conformal field theory. The first two
theories emerged from a critical point in the parameter space of Topologically Massive Gravity (TMG)\footnote{However, the sign conventions of Deser et. al. \cite{Deser1} will be disregarded in favor of that of ordinary Einstein gravity.}.
\ \\
\ \\
TMG was constructed by Deser, Jackiw and Templeton
\cite{Deser1}. Though retaining all solutions of Einstein gravity, like the BTZ black hole \cite{btz}, TMG allows for new solutions. 
Around an Anti-de Sitter background, it has propagating (massive) gravitons. For generic values in its parameter space, these modes have negative energy. Consequently, the AdS$_{3}$ background is unstable.
\ \\
\ \\
It was argued in \cite{andy1} that at the critical or ``chiral point,'' these energies vanish and that the resulting
theory is stable. Depending on the asymptotic boundary conditions for the metric two theories emerged: Chiral Gravity and Log Gravity.
\ \\
\ \\
The goal of this paper is to construct the supersymmetric extension
of Chiral Gravity, while a detailed study of the supersymmetric
extension of Log Gravity is work in progress \cite{inpr}. We focus on the simplest supersymmetric extension of TMG, Topologically Massive
Supergravity (TMSG), constructed by Deser and Kay \cite{Deser2}
and cosmologically extended by Deser in \cite{Deser3}. As in simple AdS$_{3}$ Supergravity \cite{Witten},
Chiral Supergravity might possess a holographic dual. Specifically, a chiral, ${\cal N}=1$
extremal\footnote{Extremal SCFT, as defined in \cite{gagu}, have no primaries other than the
identity of dimension less than $k^*/2$, with central charge $c=12k^*$.} two dimensional superconformal
field theory.
\ \\
\ \\
There are several reasons which motivated us to construct the supersymmetric extension of Chiral Gravity:
\ \\
\ \\
(1) Positive Energy Theorem
\ \\
\ \\
The Hamiltonian of supersymmetric theories is expressed as a
sum of squares of supercharges, $H=\Sigma~ Q^2$. Na\"{i}vely, this would suggest positive energy for TMSG for generic points in parameter space. However, in supersymmetric theories with higher derivative
interactions, the total energy does not have to be
positive, as became evident from the literature of the late 1970s.
At the time, the positive energy theorem of general relativity
\cite{wi,scya} had still not been proven. Supersymmetry seemed an
interesting path to follow, and Deser and Teitelboim
\cite{detei} showed positivity of the total energy in simple
Supergravity using $H=\Sigma Q^2$. 
\ \\
\ \\
Abbott and Deser \cite{abde} extended this proof to
Supergravity with a cosmological constant. This lead to the
idea \cite{gr} of using properties of supersymmetric theories to
understand bosonic supersymmetrizable theories by setting fermion
fields to zero, and culminated in Schoen and Yau's proof of the positive energy
theorem \cite{scya}. Techniques from Supergravity then inspired Witten's new proof \cite{wi}. 
\ \\
\ \\
It was later realized that supersymmetric higher derivative theories are more complicated.
In the $R + R^{2}$ case, the total energy is not positive definite, due
to the presence of ghosts \cite{bodest}.
\ \\
To illustrate this idea, consider a theory containing two free chiral supermultiplets. The total Lagrangian of this theory is
\begin{equation*}
\mathcal{L}=\frac{1}{2}\left(\phi_{1}^{*} \Box \phi_{1}+i\bar{\lambda}_{1} {\not}\partial \lambda_{1} - \phi_{2}^{*} \Box \phi_{2}-i\bar{\lambda}_{2} {\not}\partial \lambda_{2}\right).
\end{equation*}
In the above equation, the second multiplet
describes a ghost. The Hamiltonian and supercharge are a difference of two positive quantities, $H=H_1-H_2$ and $\mathcal{Q} = \mathcal{Q}_{1}- \mathcal{Q}_{2}$. This implies that $H={\rm tr} {\cal Q}^2=H_1-H_2$ is
not positive even though the theory is supersymmetric. In short, higher derivative supersymmetric theories may contain ghosts. These can lead to negative enregy. The positivity of the energy depends on the concrete model at hand.
\ \\
\ \\
It is an important question to ask if a positive energy theorem
for TMSG can be derived. This issue has been recently explored in
\cite{Sezgin1, Sezgin2} for the non-linear theory where the
positivity of the energy could not be shown, but a lower bound was 
derived. The puzzle of energy positivity was not solved but
only transformed into a new question, the type of solutions admitted
by the equations of motion. Our own calculations for the 
theory indicate that at the linearized level, the supersymmetric
theory mimics the bosonic theory considered in \cite{andy1}: For
generic points in parameter space, there is a massive gravity supermultiplet
with negative energy, and positivity of the total energy is not
guaranteed. However, at the chiral point the energy contribution of this multiplet vanishes, and
the total energy is positive. Thus, a consistent supersymmetric
extension of Chiral Gravity--- Chiral Supergravity--- exists.
The result of our energy calculation matches with the
recent work of Andrade and Marlof \cite{anma}, where it was shown that TMG has ghosts for
generic values of the couplings (except at the chiral point).
\ \\
\ \\
(2) Uniqueness of String Theory
\ \\
\ \\
Having found a supersymmetric generalization of Chiral Gravity one may wonder if it could be embedded in String Theory. We will
make some observations regarding the relation to String Theory in
our conclusions and will leave the detailed check of whether Chiral
Supergravity can be related to String Theory for future work.
\ \\
\ \\
(3) Extremal conformal field theories.
\ \\
\ \\
Chiral Supergravity may have an interesting dual extremal SCFT description along the lines of \cite{Witten, gagu}.
\ \\
\ \\
This paper is organized as follows: In Section 2, we describe ${\cal
N}=1$ Topologically Massive Supergravity (TMSG). We discuss in some
detail the different possibilities of having ${\cal N}=1$
supersymmetry either in the left-moving sector, the right-moving
sector, or both. In Section 3, we describe the isometries and
supersymmetry properties of the $AdS_3$ classical background we are
interested in. In Section 4, we describe the linearized theory. In
Section 4.1, we derive the linearized equations of motion. In Section
4.2, we review the graviton excitations that solve these equations
\cite{andy1} and in Section 4.3, we compute the explicit form of the
gravitini excitations, their wave functions, and conformal weights.
In Section 5, we calculate the energy and supercharge of TMSG using
the Abbott-Deser-Tekin approach \cite{abde, dete2}. Positivity of
the total energy indicates that TMSG is only stable at the chiral
point, $\mu \ell =1$, even though the theory is supersymmetric. We
finish in Section 6 with some conclusions, some comments about Log
Supergravity and an outlook.

\section{${\cal N}=1$ Topologically Massive Supergravity}
\label{TMSG section}

Shortly after the appearance of TMG \cite{Deser1} in the mid-1980s,
Deser and Kay constructed its ${\cal N}=(1,0)$ extension,
Topologically Massive Supergravity \cite{Deser2} by the
addition of a single Majorana gravitino; Deser \cite{Deser3}
provided the cosmological extension of the theory. The
underlying AdS symmetry group $SO(2,2)\approx
SL(2,\mathbb{R})_L\times SL(2,\mathbb{R})_R$ is then enhanced to
$Osp(1|2;\mathbb{R})_L\times SL(2,\mathbb{R})_R$. It is also
possible to include gravitinos in the right moving sector or in
both the right and the left moving sector. We will first present
the ${\cal N}=(1,0)$ theory and discuss the other two cases after
that. The action describing the ${\cal N}=(1,0)$ theory
is
\begin{equation}
\label{1}
\begin{split}
\mathcal{S}&=\frac{1}{16\pi G}\int d^{3}x
e\mathcal{L}=\frac{1}{16\pi G } \int d^{3}x e\left[R - 2\Lambda -
\frac{1}{2\mu}\epsilon^{\mu\nu\rho}
\left(\partial_{\mu}\omega_{\nu}^{ab}\omega_{\rho ba} + \frac{2}{3}
\omega_{\mu}^{a}{}_{b}\omega_{\nu}^{b}{}_{c
}\omega_{\rho}^{c}{}_{a}\right)\right.\\
&\hphantom{=\frac{1}{16\pi G}\int d^{3}x e\mathcal{L}=\frac{1}{16\pi
G }\int d^{3}x
e[}\left.-i\epsilon^{\mu\nu\rho}\bar{\boldsymbol{\psi}}_{\mu}\left(D_{\nu}
- \frac{1}{2\ell}\gamma_{\nu}\right)\boldsymbol{\psi}_{\rho} +
\frac{i}{2\mu}\bar{f}^{\mu}\gamma_{\nu}\gamma_{\mu}f^{\nu}\right]\;.
\end{split}
\end{equation}\label{action}
Here, $\Lambda = -\frac{1}{\ell^2}$ is the cosmological constant, the
gravitino mass parameter is equal to the reciprocal of the AdS
radius $\ell$, and $G$ is the three-dimensional Newton's constant.
The $f^{\mu}$ appearing in the action is the dual of the gravitino
field strength given by
\begin{equation}
f^{\mu}=\epsilon^{\mu\alpha\beta}D_{\alpha}\boldsymbol{\psi}_{\beta}\;;
\quad \hbox{where}\quad
D_{\alpha}\boldsymbol{\psi}_{\beta}=\partial_{\alpha}\boldsymbol{\psi}_{\beta}+
\frac{1}{4}\omega_{\alpha}^{ab}\gamma_{ab}{\boldsymbol{\psi}}_{\beta}-
\Gamma_{\alpha\beta}^{\lambda}\boldsymbol{\psi}_{\lambda}\;.
\end{equation}
In this expression (as well as in all other expressions defining the
theory) the spin connection involves torsion. We work in the second-order
formalism and define the functional form of
$\omega_{\mu}^{ab}(e,\boldsymbol{\psi})$ to be precisely that of
simple supergravity. This can be determined using the Palatini
formalism (see \cite{VanNieu}) giving
\begin{equation}
\label{ex} \omega_{\mu ab}\left(e,\boldsymbol{\psi}\right) =
\omega_{\mu ab}\left(e\right)+\kappa_{\mu ab}\left(e,\boldsymbol{\psi}\right)\;,
\end{equation}
where
\begin{equation}
\kappa_{\mu ab}\left(e,\boldsymbol{\psi}\right)=\frac{i}{4}
\left(\bar{\boldsymbol{\psi}}_{\mu}\gamma_{a}\boldsymbol{\psi}_{b}-
\bar{\boldsymbol{\psi}}_{\mu}\gamma_{b}\boldsymbol{\psi}_{a}+
\bar{\boldsymbol{\psi}}_{a}\gamma_{\mu}\boldsymbol{\psi}_{b}\right)\;.
\end{equation}
We observe that torsion comes from gravitini, and we will refer to
the connection $\omega\left(e,\boldsymbol{\psi}\right)$ as the
torsional spin connection as opposed to the standard spin
connection $\omega\left(e\right)$ defined by the vielbein
postulate
\begin{equation}
D_{\mu}e_{\nu}^{a} =
\partial_{\mu}e_{\nu}^{a}-e_{\rho}^{a}\Gamma_{\mu\nu}^{\rho}\left(g\right)
+ \omega_{\mu}^{ab}\left(e\right) e_{\nu b}=0\;,
\end{equation}
with $\Gamma_{\mu\nu}^\rho(g)$ representing the standard Christoffel
connection.
Note that the torsional spin connection also
satisfies the torsional constraint
\begin{equation}
D_{\mu} e_{\nu}^{a}=
\partial_{\mu}e_{\nu}^{a}-\Gamma_{\mu\nu}^{\rho}\left(g,\boldsymbol{\psi}\right)e_{\rho}^{a}+
\omega_{\mu}^{ab}\left(g,\boldsymbol{\psi}\right)e_{\nu b}=0\;,
 \end{equation}
which also serves as the definition of the torsional Christoffel
connection.
\ \\
\ \\
The ${\cal N}=(1,0)$ is invariant under the local supersymmetry
transformations
\begin{align}
\delta e_\mu^a &= \bar{\epsilon}\gamma^a\boldsymbol{\psi}_{\mu}\;,\\
\delta \boldsymbol{\psi}_\mu &= 2D_{\mu}\epsilon - \frac{1}{\ell}
\gamma_{\mu}\epsilon\;,
\end{align}
where $D_{\mu}\epsilon= \partial_{\mu}\epsilon +
\frac{1}{4}\omega_{\mu}^{ab}\gamma_{ab}\epsilon$ is the standard
covariant derivative of a spinor. The conformal (topological) part
of the action is separately invariant under supersymmetry.
\ \\
\ \\
The non-linear field equations for the graviton and gravitino
following from the action \eqn{1} are
\begin{align}
\mathcal{G}_{\mu\nu} + \frac{1}{\mu}C_{\mu\nu}
+F_{\mu\nu}&=0,\hbox{ and}\label{boson
field equation}\\
f^\mu
-\frac{1}{2\ell}\gamma^{\mu\nu}\boldsymbol{\psi}_{\nu}+\frac{1}{\mu}C^\mu&=0\;.\label{fermion
field equation}
\end{align}
In these equations, $\mathcal{G}_{\mu \nu}$ is the cosmologically
modified Einstein tensor,
\begin{equation}
\mathcal{G}_{\mu\nu} = R_{\mu\nu}-\frac{1}{2}g_{\mu\nu} R+\Lambda
g_{\mu\nu}\;.
\end{equation}
$C_{\mu \nu}$ is the Cotton tensor,
\begin{equation}
C_{\mu\nu} =\epsilon_{\mu}{}^{\rho\sigma}
\nabla_{\rho}\left(R_{\sigma\nu} -\frac{1}{4}g_{\sigma\nu} R\right)\;.
\end{equation}
The covariant derivative appearing
in this expression is written in terms of the torsionful Christoffel
connection. Furthermore, we denote by $F_{\mu\nu}$ the fermionic (up
to 6th order in fermions) contribution to the graviton field
equation and will give its explicit expression to the relevant order
when needed. The supersymmetric partner of the Cotton tensor, the
Cottino, is the vector-spinor
\begin{equation}
C^{\mu} = \frac{1}{2}\gamma^{\rho}\gamma^{\mu\nu}D_{\nu}f_{\rho} -
\frac{1}{8}\epsilon^{\lambda\nu\rho}R_{\lambda\nu
ab}\left(2\delta_{\rho}^{\mu}\gamma^{b}\boldsymbol{\psi}^{a}+e^{\mu
b}\gamma_{\rho}\boldsymbol{\psi}^{a}\right)\;.
\end{equation}
Having the formulas for the ${\cal N}=(1,0)$ theory, it is an easy
matter to describe the ${\cal N}=(0,1)$ theory taking into account
that parity relates both theories. A detailed discussion of the
action of parity on the ${\cal N}=(1,0)$ theory is presented in
Appendix \ref{discrete symmetries subappendix}. The action of parity effectively reduces to a sign
reversal in $\mu$ and $\ell$, so that all of the previous formulas
apply for the ${\cal N}=(0,1)$ theory after the corresponding sign
changes.
\ \\
\ \\
The action for the ${\cal N}=(1,1)$ model incorporates two Majorana
gravitinos, $\boldsymbol{\psi}_L$, $\boldsymbol{\psi}_R$ with mass
terms of opposite sign. Such an action does not seem to have been
discussed in the literature, as far as we know. However, for the
purpose of studying the linearized theory, we can easily extend the
previous formulas to a theory with $\mathcal{N}=(1,1)$ supersymmetry
by including an extra gravitino with opposite mass term into the
action \eqn{1}. In this case we can apply the Palatini formalism to
determine the torsional spin connection to be
\begin{equation}
\omega_{\mu}^{ab}(e,\boldsymbol{\psi}^{R},\boldsymbol{\psi}^{L}) =
\omega_{\mu}^{ab}(e) + \kappa_{\mu}^{ab} (\boldsymbol{\psi}^{L}) +
\kappa_{\mu}^{ab}(\boldsymbol{\psi}^{R})\;,
\end{equation}
where $\omega(e)$ and $\kappa$ are the torsion-free spin connection
and the contorsion as previously defined. Eventual interactions
between left and right gravitino fields due to this torsional
coupling
 would show up at fourth and higher orders in the perturbative expansion, but
 they are irrelevant for the
 linearized theory we are interested in.
The supersymmetry transformations for the ${\cal N}=(1,1)$ theory
are
\begin{align}
\delta e^{a}_{\mu} &=
\bar{\epsilon}\gamma^{a}\boldsymbol{\psi}^L_{\mu} - \bar{\epsilon}
\gamma^{a}\boldsymbol{\psi}^{R}_{\mu}\;,\\
\delta \boldsymbol{\psi}^L_{\mu} &= 2D_{\mu}\epsilon - {1\over
\ell}\gamma_{\mu}\epsilon\;,\\
\delta \boldsymbol{\psi}^{R}_{\mu} &= 2D_{\mu}\epsilon + {1\over
\ell}\gamma_{\mu}\epsilon\;.
\end{align}
In Section \ref{energy and supercharge section}, we calculate the energy and supercharge of the three
${\cal N}=1$ models we just discussed.

\section{The Classical Background}
\label{classical background section} 
Topological massive
supergravity has an ${\cal N}=1$ supersymmetric AdS$_3$ vacuum for
which the metric in global coordinates takes the form
\begin{equation}
ds^{2}=\bar{g}_{\mu\nu} dx^{\mu}dx^{\nu}=\ell^{2}\left(-\cosh^2\rho
d\tau^2+ \sinh^2\rho d\phi^{2}+d\rho^{2}\right)
\end{equation}
while the gravitino vanishes. In this section, we describe the form
of the (super)symmetry generators that will be used later to
calculate
 the explicit form of the bosonic and fermionic wave functions along the
lines of \cite{andy1}.

\subsection{Isometries}
\label{isometries subsection} 
AdS$_3$ is maximally symmetric and
thus has six Killing vectors, $K^{\mu}$, which generate the
$SO(2,2)\approx SL(2,\mathbb{R})\times SL(2,\mathbb{R})$ group of
isometries. When acting on scalars, their generators take the forms
\cite{andy1}:
\begin{align}
L_{0} &= K^{\mu}_{(0)}\partial_{\mu} =
\frac{i}{2}\left(\partial_{\tau} +
 \partial_{\phi}\right)\;, \label{kill0}\\
 \bar{L}_{0} &= \bar{K}^{\mu}_{(0)}\partial_{\mu} =
\frac{i}{2}\left(\partial_{\tau} -
 \partial_{\phi}\right)\;, \\
L_{\pm1} &=K^{\mu}_{(\pm 1)}\partial_{\mu}= \frac{i}{2}e^{\pm i(\tau+
\phi)} \left(\tanh\rho\partial_{\tau} + \coth\rho\partial_{\phi} \mp
i\partial_{\rho}\right)\;,\\
\bar{L}_{\pm1} &= {\bar
K}^{\mu}_{(\pm1)}\partial_{\mu}=\frac{i}{2}e^{\pm i(\tau - \phi)}
\left(\tanh\rho\partial_{\tau} - \coth\rho\partial_{\phi} \pm
i\partial_{\rho}\right)\;, \label{kill3}
\end{align}
 where unbarred and barred operators refer to the left- and
 right-moving algebras respectively.
These generators satisfy the conformal algebra
\begin{equation}
\left[L_0,L_{\pm 1}\right]=\pm L_{\pm 1}\;; \qquad
\left[L_1,L_{-1}\right]=2L_0\;,
\end{equation}
and similarly for the right moving operators. In the supersymmetric
case, we are interested in this algebra as extended to a
super-Virasoro algebra, which we elaborate on in the next Section \ref{supersymmetry subsection}.
\ \\
\ \\
The conformal algebra in the bulk is enhanced to an infinite
dimensional Virasoro algebra on the boundary \cite{henneaux}:
\begin{equation}
\left[L_{n},L_{m}\right]=\left(m-n\right)L_{m+n}+\frac{1}{12}c\left(m^3-m\right)\delta_{m+n,0}\;,
\end{equation}
and similarly for the barred algebra.
\ \\
\ \\
 It was shown in \cite{andy1} that the $SL(2,\mathbb{R})$
 algebra can be used to classify the states that satisfy the
three-dimensional
 equations of motion.\footnote{Although not all solutions
 can be obtained in this way, the logarithmic mode is a counterexample.}
 Gravitons, $|h\rangle$, are described as
primary states of this
 algebra and are labeled by the weights $(h,\bar h)$
\begin{equation}
L_0|h\rangle=h|h\rangle\;,\quad
\bar{L}_{0}|\bar{h}\rangle=\bar{h}|\bar{h}\rangle\;, \label{p1}
\end{equation}
and satisfy
\begin{equation}
L_{n}|h\rangle=\bar{L}_{n}|\bar{h}\rangle=0\quad n>0\;.\label{p2}
\end{equation}
Equivalently, we can label these states by their energy
$E=h+\bar{h}$ and their spin $S=h-\bar{h}$. Unitarity of the
representation (e.g. positivity of the norm of all states) imposes
constraints on the central change and weight of the primary fields
\cite{Fri,God}. Unitary representations exist for all values $(c,h)$
with $c \geq 1$ and $h\geq 0$, or equivalently
$E\geq\left|S\right|$. Representations that saturate this bound are
called ``massless'' and describe non-propagating degrees of
freedom. Representations with $E>\left|S\right|$ are called
``massive'' and describe propagating states with helicity $S$. As
will be checked later, primary states in both representation appear
in TMSG.
\ \\
\ \\
It was shown by Brown and Henneaux \cite{henneaux} that for ordinary
Einstein theory, the central charges of the left- and right-moving
algebra are equal:
\begin{equation}
c_{L} = c_{R} = \frac{3\ell}{2G}\;.
\end{equation}
The gravitational Chern-Simons term appearing in TMG deforms these
central charges so they are no longer equal
\cite{KrausandLarsen}
\begin{equation}
\label{central charges} c_{L} = \frac{3\ell}{2G}\left(1 -
\frac{1}{\mu\ell}\right), \quad c_{R} =\frac{3\ell}{2G} \left(1 +
\frac{1}{\mu\ell}\right)\;.
 \end{equation}
Taking into account that the semiclassical approximation corresponds
to large $\ell/G$, it is interesting to note that unitarity of the
boundary theory demands $|\mu \ell| \geq 1$. For the specific
value of the Chern-Simons coupling, ${\mu} \ell= 1$, it was noticed
in \cite{andy1} that $c_{L} = 0$ and $c_{R} = 3\ell/G$. This
suggests that right-moving gravitational physics may behave as in
pure gravity (albeit with twice the
  central charge) and that left-moving gravitational physics might be trivial.
  In other work, it was shown that once particular boundary conditions
  are imposed, it is indeed possible to obtain a left-moving theory that is
trivial
  \cite{Maloney, grjo1,  CarlipDeser1,  CarlipDeser2, porrati}.

\subsection{Supersymmetry}
\label{supersymmetry subsection}
We consider again all three cases, i.e. supersymmetry in the left
sector, the right sector or both.
 For the ${\cal N}=(1,0)$ theory
the Killing spinor equation takes the form
\begin{equation}
\delta \boldsymbol{\psi}_{\mu} = 2D_{\mu}\epsilon -
\frac{1}{\ell}\gamma_{\mu}\epsilon =  0\;,
\end{equation}
where $\gamma_{\mu}=e^a_{\mu}{\gamma}_a$. There are two Killing spinors
that solve this equation:
\begin{equation}
\xi_{L} =
e^{(iu-\rho)/2}\begin{pmatrix}-ie^{\rho}\cr 1\end{pmatrix}\quad\hbox{and}
 \quad \xi_{L}^{*} =
 e^{-(iu+\rho)/2}\begin{pmatrix}ie^{\rho}\cr 1\end{pmatrix}\;,
\end{equation}
where $u = \tau + \phi$. Like the Killing vectors $K_{1}^{\mu}$ and
$K_{-1}^{\mu}$, these Killing spinors are simply complex conjugates
of one another. Killing spinors are associated with the fermionic
generators of the super-Virasoro algebra. In general, fermionic
fields depending on a compact coordinate (in this case the $\phi$
coordinate) can be either periodic or anti-periodic under $\phi
\rightarrow
 \phi + 2\pi$, corresponding either the Ramond (R) or Neveu-Schwarz (NS)
sector algebra.
 Since the above spinors are anti-periodic, we are interested in the NS
sector of the algebra.
 In this sector, the $\mathcal{N} = (1,0)$ global subalgebra is generated by
\begin{equation}
L_{0},L_{\pm 1},G_{\pm 1/2},\bar{L}_{0},\bar{L}_{\pm 1}.
\end{equation}
The left-moving superconformal algebra is
\begin{equation}
\left[L_{m},L_{n}\right]=\left(m-n\right)L_{m+n}\;,\quad
\left[L_{m},G_{r}\right]= \left(\frac{m}{2}-r\right)G_{m+r}\;,\quad
\left\{G_{r},G_{s}\right\}=2L_{r+s}\;,
\end{equation}
with $m,n=0,\pm 1$ and $r,s=\pm 1/2$. Similarly, as in the bosonic
case, we expect this algebra to be enhanced to an infinite
dimensional super-Virasoro algebra on the boundary
\begin{align}
\left[L_{m},L_{n}\right]&=\left(m-n\right)L_{m+n}+\frac{1}{12}c_L\left(m^3-m\right)\delta_{m+n,0}\;,\\
\left\{G_{m},G_{n}\right\}&=2L_{m+n}+\frac{1}{3}c_L\left(m^2-\frac{1}{4}\right)\delta_{m+n,0}\;,\\
\left[L_{m},G_{n}\right]&=\left(\frac{m}{2}-n\right)G_{m+n}\;.
\end{align}
The central change of the supersymmetric theory is the same as the
central charge of the
 bosonic theory (see \cite{bbcho}).
 \ \\
\ \\
In the supersymmetric case graviton and gravitinos are primary
fields of the super-Virasoro algebra that satisfy the constraints
\eqn{p1}, \eqn{p2} and
\begin{equation}
G_{r}|h\rangle=0\;, \quad r >0\;.\label{p3}
\end{equation}
Although the explicit form of the fermionic generators will not be
needed to obtain the graviton and gravitino wave functions, notice
that just as $L_{m}$ can be expressed in terms of Killing
vectors, the fermionic
 generators, $G_{\pm 1/2}$, can be expressed in terms of Killing
spinors. For
  example, the action of $L_{1}$ on a scalar field $\phi$ is simply the
directional derivative
\begin{equation}
L_{1}\phi = K^{\mu}_{1}\partial_{\mu}\phi\;.
\end{equation}
Similarly, the action of $G_{1/2}$ on $\phi$ is
\begin{equation}
G_{1/2}\phi = \gamma^{\mu}\xi_{L}\partial_{\mu}\phi\;.
\end{equation}
The $\mathcal{N}=(0,1)$ theory involves an
inequivalent representation of the Clifford algebra, see Appendix 
\ref{clifford algebra subappendix} for details. A study of the $\mathcal{N}=\left(0,1\right)$ yields the resulting Killing spinor equation:
\begin{equation}
D_{\mu}\epsilon + \frac{1}{2\ell}\gamma_{\mu}\epsilon = 0\;.
\end{equation}
There are again two Killing spinors that solve this equation that
take the form
\begin{equation}
\xi_{R} = e^{(iv - \rho)/2}\begin{pmatrix}i\cr e^{\rho}
\end{pmatrix}, \quad \xi^{*}_{R} = e^{-(iv + \rho)/2}
\begin{pmatrix}-i\cr e^{\rho}\end{pmatrix}\;.
\end{equation}
Here, $v = \tau - \phi$, and we observe again that both spinors are
complex conjugates of one another.
\ \\
\ \\
Thus, we see that $AdS_{3}$ is a viable supersymmetric 
background for an $\mathcal{N}=\left(1,1\right)$ extension of TMG.
\ \\
\ \\
Next, we analyze the stability of the $AdS_3$ background by
considering perturbations around it.
 We proceed as in \cite{andy1}: Compute the explicit form of linear graviton
  and gravitino fields, then compute conserved charges to second order in
these
  perturbations. First, however, we must discuss some generalities.

\section{The Perturbative Expansion}
\label{perturbative expansion section}

Due to the nonlinearity present in Topologically Massive Gravity and
the corresponding supergravity theories, one is often forced into a
perturbative regime as to make headway. Indeed, the analysis
leading to Chiral Gravity and our corresponding work on
Topologically Massive Supergravity relies heavily on
 the methods of perturbation theory. In this section, we will briefly
recapitulate
 the basics of perturbative gravity, and then proceed to determine the
perturbative
  spectrum of the supersymmetric theory.
\ \\
\ \\
Expansion in small fluctuations about some fundamental object is central to all variants of perturbation theory. In gravitational theories, the
fundamental
  field is the metric, the starting point of perturbative
gravity is to expand
   in fluctuations about some background metric, $\bar{g}_{\mu\nu}$, which
is a known solution
    of the theory, as follows
\begin{equation}
g_{\mu\nu}=\bar{g}_{\mu\nu}+\lambda
h_{\mu\nu}+\lambda^{2}j_{\mu\nu}+\mathcal{O}
\left(\lambda^{3}\right)\;.\label{metric perturbation}
\end{equation}
Here, $\lambda$ is a small parameter to be used as a book-keeping
device. Thus the $n$$^{\rm{th}}$ order in perturbation theory is
tantamount to $\mathcal{O} \left(\lambda^{n}\right)$ in the relevant
expansions. In supergravity theories there is a second field, the
gravitino, which should also be expanded in powers of $\lambda$. It is important to note that the gravitino
is a fermion and so when performing such an expansion the background term is identically zero:
\begin{equation}
\boldsymbol{\psi}_{\mu}=\lambda\psi_{\mu}+\lambda^{2}\psi^{(2)}_{\mu}+\mathcal{O}
\left(\lambda^{3}\right)\;.\label{rarita schwinger perturbation}
\end{equation}
Given such perturbative expansions for the fundamental fields of the
theory, one constructs perturbative expansions for all
objects appearing in the action. In general, given
 a multilinear map $M_{\mu\nu...}\left(g,\boldsymbol{\psi}\right)$, its
formal perturbative
  expansion can be written as
\begin{equation}
M_{\mu\nu\ldots}\left(g,\boldsymbol{\psi}\right)=M^{(0)}_{\mu\nu\ldots}+
\lambda M^{(1)}_{\mu\nu\ldots}+\lambda^{2}
M^{(2)}_{\mu\nu\ldots}+\mathcal{O}
\left(\lambda^{3}\right)\;,\label{general tensor perturbative
expansion}
\end{equation}
where the functional form of $M^{(n)}_{\mu\nu\ldots}$ is given by
\begin{equation}
M^{(n)}_{\mu\nu\ldots}=\frac{1}{n!}\left.\frac{\partial^{n}M_{\mu\nu\ldots}
\left(g,\boldsymbol{\psi}\right)}{\partial\lambda^{n}}\right|_{\lambda=0}\;.
\end{equation}
Applying such an expansion to the equations of motion allows one to
work order
 by order in $\lambda$ and generate the perturbative spectrum of the theory.
 
 \subsection{Linearized Supergravity}
\label{Linearized SUGRA}

Consider first the ${\cal N}=(1,0)$ theory. In the linearized
approximation the field equations of this theory take the form
\begin{equation}
\mathcal{G}^{(1)}_{\mu\nu}\left(h\right)+\frac{1}{\mu}C^{(1)}_{\mu\nu}
\left(h\right)=0\;,
\end{equation}
and
\begin{equation}
f^{(1)}_{\mu}\left(\psi\right)-\frac{1}{2\ell}\bar{\gamma}_{\mu\nu}\psi^{\nu}
+\frac{1}{\mu}C_{\mu}^{(1)}\left(\psi\right)=0\;,
\end{equation}\label{le}
where barred quantities are expressed with respect to the background metric.
Here we have introduced the notation
\begin{align}
\mathcal{G}^{(1)}_{\mu\nu}\left(h\right)&=R^{(1)}_{\mu\nu}\left(h\right)-
\frac{1}{2}\bar{g}_{\mu\nu}R^{(1)}\left(h\right)-2\Lambda
h_{\mu\nu}\;,
 \label{linearized einstein tensor}\\
C^{(1)}_{\mu\nu}\left(h\right)&=\epsilon^{\alpha\beta}{}_{\mu}
\bar{\nabla}_{\alpha}\left(R^{(1)}_{\beta\nu}\left(h\right)-\frac{1}{4}
\bar{g}_{\beta\nu}R^{(1)}\left(h\right)-2\Lambda
h_{\beta\nu}\right)\;, \label{linearized cotton tensor}
\end{align}
with
\begin{align}
R^{(1)}(h)&=-\bar{\nabla}^{2} h +
\bar{\nabla}_{\mu}\bar{\nabla}_{\nu}h^{\mu\nu} - 2\Lambda h\;,\\
R^{(1)}_{\mu\nu}\left(h\right)&=\frac{1}{2}\left(-\bar{\nabla}^{2}
h_{\mu\nu}-\bar{\nabla}_{\mu}\bar{\nabla}_{\nu}h+\bar{\nabla}^{\sigma}
\bar{\nabla}_{\nu}h_{\sigma\mu}+\bar{\nabla}^{\sigma}\bar{\nabla}_{\mu}h_{\sigma\nu}\right)\;,\\
f^{(1)}_{\mu}\left(\psi\right)&=\epsilon_{\mu\alpha\beta}\bar{D}^{\alpha}\psi^{\beta}\;,\\
C^{(1)\mu}\left(\psi\right)&=\frac{1}{2}\bar{\gamma}^{\rho}
\bar{\gamma}^{\mu\nu}\bar{D}_{\mu}f^{(1)}_{\rho}\left(\psi\right)-\frac{1}{4\mu\ell^{2}}\psi_{\mu}.
\end{align}
Note that at linear order the bosonic and
fermionic
 equations completely decouple. This is due to gravitons
 and gravitini coupling through torsion, which is a second order effect in
$\lambda$. This
 allows us to study the graviton wave functions separately from the
gravitinos. The form of
 the linearized bosonic equation of motion is precisely
the one found by  \cite{andy1}, and so before proceeding to the case
of
  the Rarita-Schwinger field, we will review the metric fluctuations.

\subsection{Linearized Gravitons}
\label{linearized gravitons subsection} 
The graviton field equation
becomes much simpler if a particular gauge is chosen. It was shown
in \cite{andy1} that a convenient gauge choice is the
divergence free-gauge. At the linear level it reads
\begin{equation}
\bar{\nabla}_{\mu}\left(h^{\mu\nu}+\bar{g}^{\mu\nu}h\right)=0\;.
\end{equation}
Combining this with the bosonic field equation yields the traceless
condition $h=0$, so
 that the linearized bosonic field can be chosen to be divergence free and
traceless:
\begin{equation}
h=0,\quad\bar{\nabla}_{\mu}h^{\mu\nu}=0\;.
\end{equation}
It was further shown in \cite{andy1} that implementation of this
gauge condition reduces the linearized bosonic equation of motion to
\begin{equation}
\left(\bar{\nabla}^{2}+\frac{2}{\ell^{2}}\right)\left(h_{\mu\nu}+
\frac{1}{\mu}\epsilon_{\mu}{}^{\rho\sigma}\bar{\nabla}_{\rho}h_{\sigma\nu}\right)=0\;.
\end{equation}
This equation can be written in terms of the
SL$\left(2,\mathbb{R}\right)$ Casimir,
$L^{2}+\bar{L}^{2}=-\frac{\ell^{2}}{2}\nabla^{2}$, which motivated
\cite{andy1} to use the SL$\left(2,\mathbb{R}\right)$ algebra to
find the solutions to the equation of motion. Gravitons are
described by harmonic functions on $AdS_{3}$, which take the form
\begin{equation}
h_{\mu\nu}=e^{-i\left(E\tau+S\phi\right)}M_{\mu\nu}\left(\rho\right)\;,
\end{equation}
where $M_{\mu\nu}$ is a symmetric two-index tensor depending only on
$\rho$ whose explicit form will be calculated. The spin is determined 
by the gauge choice for $h_{\mu\nu}$ and is found to be $S=\pm2$.
Furthermore, $M_{\mu\nu}\left(\rho\right)$ can be determined through
the application of
 the primary field constraints
\begin{equation}
L_{1}h_{\mu\nu}=\bar{L}_{1}h_{\mu\nu}=0\;.
\end{equation}
These constraints can be rearranged into the more convenient form
\begin{equation}
\left(L_{1}\pm\bar{L}_{1}\right)h_{\mu\nu}=0\;.\label{boson primary field constraint}
\end{equation}
The generators, $L_{n}$, ${\bar L}_{n}$ are taken to be Lie
derivatives along the Killing vector fields. In particular, when 
acting on a two-index tensor, the Lie
derivative along a vector field $K_{(n)}$ takes the form
\cite{ortin}:
\begin{equation}
L_{n}h_{\mu\nu}=K^{\lambda}_{(n)}\left(\bar
\nabla_{\lambda}h_{\mu\nu}\right)+ \left(\bar
\nabla^{\lambda}K_{(n)\mu}\right)h_{\lambda\nu}+\left(\bar
\nabla^{\lambda}K_{(n)\nu}\right) h_{\mu\lambda}\;.
\end{equation}
The f fields in these equations are taken to be the Killing
 vectors $K_{(n)}^{\mu}$, ${\bar K}_{(n)}^{\mu}$ defined in
\eqn{kill0}-\eqn{kill3}.
  With the lower sign of \eqn{boson primary field constraint}, the
tensor $M_{\mu\nu}$ can be determined to be
\begin{equation}
M_{\mu\nu}\left(\rho\right)=f\left(\rho\right)\begin{pmatrix}
1
 & \frac{S}{2} & ia \cr \frac{S}{2} & 1 & \frac{iSa}{2} \cr ia & \frac{iSa}{2}
  & -a^{2}\end{pmatrix}\;,
\end{equation}
with
\begin{equation}
a=\frac{1}{\sinh\rho\cosh\rho}.
\end{equation}
Taking the upper sign of \eqn{boson primary field constraint} allows
one to determine the matrix prefactor, $f\left(\rho\right)$. In
particular, one arrives at the differential equation
\begin{equation}
\partial_{\rho}f\left(\rho\right)+\frac{E\sinh^{2}\rho-2\cosh^{2}\rho}
{\sinh\rho\cosh\rho}f\left(\rho\right)=0\;,
\end{equation}
which admits the solution
\begin{equation}
f\left(\rho\right)=\frac{\sinh^{2}\rho}{\cosh^{E}\rho}\;.
\end{equation}
Combining these results yields the graviton wave function up to
overall
 normalization \cite{andy1}:
\begin{equation}
h_{\mu\nu}=N_{b}e^{-i\left(E\tau+S\phi\right)}\frac{\sinh^{2}\rho}
{\cosh^{E}\rho}\begin{pmatrix} 1 & \frac{S}{2} & ia \cr
\frac{S}{2} &
 1 & \frac{iSa}{2} \cr ia & \frac{iSa}{2} & -a^{2}\end{pmatrix}\;.
\end{equation}
The energy value $E$ can be determined by inserting this
solution into the linearized equation of motion. Upon
 restricting to normalizable modes, the weights or energy and
spin
 can be fixed to
\begin{equation}
\left(E,S\right)=\left(2,\pm2\right)\quad \hbox{or}\quad
\left(1\pm\mu\ell,\pm2\right). \label{boson mode weights}
\end{equation}
In \cite{andy1}, $\left(E,S\right)=\left(2,2\right)$ is referred to
 as the left-moving graviton,
$\left(E,S\right)=\left(2,-2\right)$ is the
 right-moving graviton, and
$\left(E,S\right)=\left(1+\mu\ell,2\right)$
 is the massive graviton. The final case,
$\left(E,S\right)=\left(1-\mu\ell,-2\right)$,
 is not considered a solution to the theory since the wave function
is non-normalizable. At the chiral point, $\mu\ell=1$, the wave
function of the massive mode coincides with the one of the left-moving graviton.
   It was argued in \cite{andy1} that for suitable boundary
 conditions this left moving wave function can be gauged away so that the
 theory becomes chiral, with only a right-moving degree of
 freedom.

\subsection{Linearized Gravitini}
\label{linearized gravitini subsection}

Deriving the gravitino wave functions at the linear level proceeds
in a similar fashion. Consider again first the ${\cal N}=(1,0)$
theory. The left-moving gravitino wave functions are vector-spinors
on $AdS_{3}$
\begin{equation}
\psi_{\mu}=e^{-i\left(E\tau+S\phi\right)}\zeta_{\mu}\left(\rho\right)\;,\label{vs}
\end{equation}
where $\zeta_{\mu}$ with $\mu=0,1,2$ is a two-component spinor
depending only on $\rho$. As with the gravitons, a suitably chosen
gauge simplifies the equations of motion.
\ \\
\ \\
It is understood that the Rarita-Schwinger field carries its own
gauge freedom. Specifically, equivalent physical states are obtained
by
$\psi_{\mu}\rightarrow\psi_{\mu}+(D_{\mu}\pm\frac{1}{2\ell}\gamma_{\mu})\kappa$,
where $\kappa$ is some spinor field. This gauge freedom allows one
to fix
\begin{equation}
\bar \gamma^{\mu}\psi_{\mu}=0\;.\label{gt}
\end{equation}
This is the natural choice for the superpartner of the traceless
graviton. In fact, applying a supersymmetry transformation to the
linearized graviton trace-free
 gauge condition yields the gamma-traceless condition \eqn{gt}. Expanding
this condition yields
 the relationship
\begin{equation}
\psi_{2}=\bar \gamma_{1}\psi_{0}-\bar \gamma_{0}\psi_{1}\;,
\end{equation}
which can be used to determine $\psi_2$ once $\psi_0$ and $\psi_1$
are known. To determine $\psi_{0}$ and $\psi_{1}$ it is sufficient
to apply the lowest-weight/primary-field conditions
\begin{equation}
\left(L_{1}\pm\bar{L}_{1}\right)\psi_{\mu}=0\;,\label{fermion primary
field constraint}
\end{equation}
where, as in the bosonic case, the $L_{1}$, ${\bar L}_1$ operators
are Lie derivatives along the Killing vector fields $K_{1}^{\mu}$,
${\bar K}_{1}^{\mu}$ acting on a vector-spinor. Specifically, they
are given by \cite{ortin}:
\begin{equation}
L_{n}\psi_{\mu}=K^{\lambda}_{(n)}\bar
D_{\lambda}\psi_{\mu}+\frac{1}{2}( \bar{\nabla}_{\alpha}
K_{(n)\beta})\bar \gamma^{\alpha\beta}\psi_{\mu}+\left(\bar
\nabla_{\mu}K_{(n)}^{\lambda}\right)\psi_{\lambda}\;.
\end{equation}
However, when $K_{(n)}^{\lambda}$, ${\bar K}_{(n)}^{\lambda}$ are
Killing vectors, one can apply the AdS$_{3}$ algebra to reduce these
expressions to
\begin{equation}
L_{n}\psi_{\mu}=K^{\lambda}_{(n)}(\bar D_{\lambda} -
\frac{1}{2\ell}\bar \gamma_{\lambda}) \psi_{\mu}+\left(\bar
\nabla_{\mu}K^{\lambda}_{(n)}\right)\psi_{\lambda}
\end{equation}
and
\begin{equation}
{\bar L}_{n}\psi_{\mu}={\bar K}^{\lambda}_{(n)}(\bar D_{\lambda} +
\frac{1}{2\ell}\bar \gamma_{\lambda}) \psi_{\mu}+\left(\bar
\nabla_{\mu}{\bar K}^{\lambda}_{(n)}\right)\psi_{\lambda}\;.
\end{equation}
Choosing the minus sign in \eqn{fermion primary field constraint},
one finds that
 $S=\frac{3}{2}$ and
\begin{equation}
\psi_{\mu}=N_{f}e^{-iE\tau-iS\phi}F_{\mu}\left(\rho\right)\begin{pmatrix}
i\cr
 e^{\rho}\end{pmatrix}\;.
\end{equation}
Here,
\begin{equation}
F_{0}\left(\rho\right)=F_{1}\left(\rho\right)=F\left(\rho\right),\quad F_{2}
\left(\rho\right)=\frac{iF\left(\rho\right)}{\sinh\rho\cosh\rho}\;,
\end{equation}
and $N_{f}$ is some overall normalization. $F\left(\rho\right)$ can
now be fixed by
 choosing the positive sign in \eqn{fermion primary field constraint},
which leads to
  the differential equation
\begin{equation}
\partial_{\rho}\psi_{1} + (E\tanh\rho -\coth\rho)\psi_{1} +
 \frac{i\bar \gamma_{1}}{2\cosh\rho}\psi_{1} = 0
\end{equation}
with solution
\begin{equation}
F\left(\rho\right)=\frac{e^{-\rho/2}\sinh\rho}{\cosh^{E+1/2}\rho}\;.
\end{equation}\label{F}
Notice that since $\psi_{\mu}$ are harmonic functions on $AdS_{3}$
they
 satisfy the Dirac equation
\begin{equation}
\left({\not}D-\frac{\left(E-1\right)}{\ell}\right)\psi_{\mu}=0\;.
\end{equation}
After fixing the gamma-traceless gauge, the linearized
fermionic equation reduces to
\begin{equation}
\left({\not}D-\frac{1}{2\ell}\right)\psi_{\mu}-\frac{1}{\mu}\left({\not}D^{2}-
\frac{1}{4\ell^{2}}\right)\psi_{\mu}=0\;,\label{linearized gravitini
eom}
\end{equation}
where the Feynman slash notation has been adopted so that
${\not}D=\bar \gamma^{\mu}D_{\mu}$. Inserting the linearized modes
into this equation yields
\begin{equation}
\frac{E-1}{\ell}\psi_{\mu}-\frac{1}{2\ell}\psi_{\mu}-\frac{1}{\mu}
\left(\frac{\left(E-1\right)^{2}}{\ell^{2}}-\frac{1}{4\ell^{2}}\right)\psi_{\mu}=0\;,
\end{equation}
which fixes the value of $E$. The energy and spin of the left-moving
gravitini are
\begin{equation}
\left(E,S\right)=\left(\frac{3}{2},\frac{3}{2}\right)\quad\hbox{or}\quad\left
(\frac{1}{2}+\mu\ell,\frac{3}{2}\right)\;.\label{w1}
\end{equation}
The former is clearly a solution to simple supergravity and hence
satisfies the requisite Dirac equation with appropriate mass. It
corresponds to the left moving gravitino, the supersymmetric partner
of the left moving graviton. The second mode corresponds to the
so-called fermionic ``massive'' propagating degree of freedom. As in
the
 bosonic theory, we observe that at the chiral point $\mu\ell=1$ the
 wave functions of the massless and massive gravitino coincide.
 As will be shown in the next section, this chiral behavior will extend
 to the conserved charges, hence Topologically Massive Supergravity
preserves
  the chiral structure found in \cite{andy1}.
\ \\
\ \\
Note that in the classical supergravity analysis, it is understood
that fermions are Grassmann-valued Majorana spinors. Thus, the
physical wave functions are the real (or, alternatively, imaginary)
parts of $\psi_{\mu}$ and there are implicit Grassmann-valued
numbers associated with all fermion spinor components. The ``physical'' temporal component of $\psi_{\mu}$ may be
 written as
\begin{equation}
\mathop{Re}\left(\psi_{0}\right)=\frac{\mathop{Re}\left(N_{f}\right)e^{-\rho/2}\sinh{\rho}}
{\cosh^{E+1/2}\rho}\begin{pmatrix}\sin\left(E\tau+S\phi\right)\theta_{1}\cr
e^{\rho}\cos\left(E\tau+S\phi\right)\theta_{2}\end{pmatrix}\;,
\end{equation}
where $\theta_{i}\theta_{j}=-\theta_{j}\theta_{i}$, and similarly for
the other
components of $\psi_{\mu}$. 
\ \\
\ \\
The previous analysis can be carried over for the ${\cal N}=(0,1)$
theory, by taking the corresponding sign changes in $\mu$ and $\ell$
into account. Given the ansatz \eqn{vs} and the primary field
constraint \eqn{fermion primary field constraint}, it is
straightforward to show that the spin is fixed to $S=-\frac{3}{2}$
and that right-moving gravitino fields are given by
\begin{equation}
\psi_{\mu} = N_{f}e^{-iE\tau
-iS\phi}F_{\mu}(\rho)\begin{pmatrix}-ie^{\rho}\cr 1
\end{pmatrix}\;,
\end{equation}
where
\begin{equation}
F_{0}(\rho) = - F_{1}(\rho) = F(\rho), \quad F_{2}(\rho) =
\frac{iF(\rho)}{\sinh\rho\cosh\rho}\;,
\end{equation}
 and $F(\rho)$ is given by the same expression as before. Note the sign
change for $F_{1}$. The equations of motion then fix the energy and
spin to be
\begin{equation}
(E,S) = (\frac{3}{2},-\frac{3}{2}) \quad \mathrm{or}\quad
(\frac{1}{2} + \mu\ell,-\frac{3}{2})\;.\label{w2}
\end{equation}
These states correspond to a right-moving gravitino and a massive
gravitino that propagates in the bulk. The ${\cal N}=(1,1)$ theory
will contain three gravitinos with conformal weights given by
\begin{equation}
\left(E,S\right)=\left(\frac{3}{2},\pm\frac{3}{2}\right)\quad\hbox{and}\quad\left(\frac{1}{2}+\mu\ell,\frac{3}{2}\right)\;.
\end{equation}
\section{Energy and Supercharge in TMSG}
\label{energy and supercharge section} 
In this section the stability
under perturbations of Topologically Massive Supergravity
 is analyzed. This is done through the study of conserved charges as
defined by Abbott and Deser
 \cite{abde}. The energy was previously
calculated for the bosonic model
  \cite{andy1} through Hamiltonian methods, where it was shown that for
generic
  values of $\mu$, the energies of some of the modes are negative,
indicating an instability
  in the theory. However, at the chiral point, the energies for all linear
perturbations
  are positive semi-definite, and so the theory defined on an $AdS_{3}$
background is stable
  against metric perturbations.
  \ \\
\ \\
We first compute the energy of individual modes and show that TMSG
on an
  $AdS_{3}$ background is unstable in general. Stability is restored at
the chiral point
  $\mu \ell =1$. One might have hoped to make
    use of supersymmetry to arrive at a positive energy
     theorem. As may be inferred, one is not able to show
     positivity. As explained in the introduction,
     this failure can be traced back to the existence of ghosts in higher
derivative theories.
     Indeed, the recent calculation of \cite{anma} shows that TMG
     has ghosts for $\mu\ell\neq1$.
     \ \\
\ \\
Once the energy is computed, we extend the analysis to determine the
supercharge. All conserved charges of the supersymmetric theory
exhibit chiral behavior allowing us to show that the theory is stable against perturbations. Thus, one can
  speak of Chiral Supergravity.

\subsection{Bosonic Charges: Energy}
\label{energy subsection} 
In their work \cite{abde} Abbott
and Deser noted that in the presence of
 a non-vanishing cosmological constant, the conventional definition of
gravitational
 energy fails and they subsequently determined a modified definition. Their key
observation
 was that for a non-vanishing cosmological constant, the space-time is not
asymptotically
 flat. This leads to the failure of conservation
  for the conventional charges. To rectify the situation, these authors
constructed conserved
  charges for a
background that satisfies Einstein's equations with arbitrary
cosmological constant
\begin{equation}
\mathcal{G}^{\mu\nu}\equiv R^{\mu\nu}-\frac{1}{2}g^{\mu\nu}R+\Lambda
g^{\mu\nu}=0\;.
\end{equation}
To construct the charges, the metric is divided into two parts: a
background value $\bar{g}_{\mu\nu}$ and a deviation $h_{\mu\nu}$,
which does not have to be small, but needs to vanish fast enough at
infinity
\begin{equation}
g_{\mu\nu}=\bar g_{\mu\nu}+h_{\mu\nu}\;.
\end{equation}
Insertion of this expansion into the Einstein
  equations allows for the definition of the energy momentum
pseudo-tensor. This
   is done by partitioning the equation of motion into three pieces: a
piece dependent only
   on the background, a piece linear in metric fluctuation, and a final
term containing
   all terms quadratic or higher order in $h_{\mu\nu}$. Since the background is Einsteinian the zeroth order contribution vanishes immediately. One then takes the
terms
    nonlinear in metric fluctuations to define the energy momentum
    pseudo-tensor $T_{\mu\nu}$,
     so that the equation of motion reads
\begin{equation}
\mathcal{G}^{(1)}_{\mu\nu}=T_{\mu\nu}
\end{equation}
where
\begin{equation}
\mathcal{G}^{(1)}_{\mu\nu}=R^{(1)}_{\mu\nu}-\frac{1}{2}\bar{g}_{\mu\nu}R^{(1)}-
\Lambda h_{\mu\nu}\;.
\end{equation}
Here, taking the superscript $(n)$ to denote the expansion of the
relevant object containing terms only of order $n$ in $h_{\mu\nu}$.
It is straightforward to show that the energy momentum pseudo-tensor
satisfies the background Bianchi identity
\begin{equation}
\bar{\nabla}_{\mu}T^{\mu\nu}=0\;.
\end{equation}
One can obtain  conserved currents by contracting the energy
 momentum pseudo-tensor with a Killing vector $\xi_{\nu}$, \cite{abde}:
\begin{equation}
\partial_{\mu}\left(T^{\mu\nu}\xi_{\nu}\right)=0\;.
\end{equation}
When the Killing vector is taken to be time like, this defines the
gravitational energy-momentum density, thus the gravitational energy
\begin{equation}
E\left(\xi_{\nu}\right)\equiv\frac{1}{8\pi G}\int ed^{3}xT^{0\nu}\xi_{\nu}\;.
\end{equation}
A similar analysis can be carried out for higher-derivative gravities (see \cite{dete1}\cite{dete2}). 
They later applied their analysis to TMG . In this case, the equations
  are modified by the presence of the Cotton tensor, and the background is now
taken to be a
   solution of the vacuum equations of TMG. Insertion of the
   expansion yields
\begin{equation}
\mathcal{G}^{(1)\mu\nu}+\frac{1}{\mu}C^{(1)\mu\nu}= T^{\mu\nu}\;.
\end{equation}
Here, $\mathcal{G}^{(1)\mu\nu}$ is the linearized Einstein tensor,
$C^{(1)\mu\nu}$ is the Cotton tensor with only terms linear in
metric fluctuations retained and the stress
energy
 pseudo-tensor is the collection of all terms quadratic or higher order in
$h_{\mu\nu}$.
 The expressions for the linearized Einstein and Cotton
 tensors were given in Section \ref{perturbative expansion section}. To obtain the explicit form of the
 energy momentum pseudo-tensor of the ${\cal N}=(1,0)$ theory, we
 apply the perturbation expansion of Section \ref{perturbative expansion section}
\begin{equation}
g_{\mu\nu}=\bar{g}_{\mu\nu}+\lambda
h_{\mu\nu}+\lambda^{2}j_{\mu\nu}+\mathcal{O}
\left(\lambda^{3}\right)\;,
\end{equation}
and
\begin{equation}
\boldsymbol{\psi}_{\mu}=\lambda\psi_{\mu}+\lambda^{2}\psi^{(2)}_{\mu}+\mathcal{O}
\left(\lambda^{3}\right)\;.
\end{equation}
We remind the reader that the gravitino expansion starts
at $\mathcal{O} \left(\lambda\right)$ due to the absence of
background fermions. Applying these expansions to the graviton field
equation \eqn{boson field equation} of TMSG and working to
$\mathcal{O}\left(\lambda^{2}\right)$, one finds
\begin{equation}
G^{(1)}_{\mu\nu}(j) + \frac{1}{\mu}C^{(1)}_{\mu\nu}(j) = -
G^{(2)}_{\mu\nu}(h,\psi) -
\frac{1}{\mu}C^{(2)}_{\mu\nu}(h,\psi)-F^{(2)}_{\mu\nu}\left(\psi\right)
= T^{(2)}_{\mu\nu}\;,
\end{equation}
where the functional forms for $\mathcal{G}^{(n)}_{\mu\nu}$ and
$C^{(n)}_{\mu\nu}$ can be determined by the general procedure
\eqn{general tensor perturbative expansion} and are
 given explicitly in Appendix \ref{notation and conventions appendix}.
Given these explicit
  forms, one can obtain the Abbott-Deser-Tekin gravitational energy to
$\mathcal{O}
  \left(\lambda^2\right)$ in the fashion discussed above. To do so, one
identifies the
  time-like Killing vector as
\begin{equation}
\xi^{\mu}=\begin{pmatrix}i\cr 0\cr 0\end{pmatrix}\;.
\end{equation}
The bosonic contributions to the energy determined in this fashion
exactly agree with the results of \cite{andy1}, the Hamiltonian
approach was used. Next, to quadratic order, the fermionic contribution to
the energy is given by
\begin{equation}
\label{fermionic energy} E_{F}=\frac{1}{8\pi
G\ell}\left(1-\frac{1}{2\mu\ell}\left(1+4\left(1-E\right)^{2}\right)
\right)\int
ed^{3}x\epsilon^{0\mu\nu}S_{\nu\mu0}^{(2)}\left(\psi\right)\;,
\end{equation}
where we have introduced the torsion, $S_{\mu\nu}{}^{\rho}$, given by the antisymmetric part of the torsional Christoffel
connection \cite{VanNieu} and explicitly by
\begin{equation}
S_{\mu\nu}{}^{\rho}=\frac{1}{4}\bar{\boldsymbol{\psi}}_{\mu}\gamma^{\rho}\boldsymbol{\psi}_{\nu}\;.
\end{equation}
As usual, the $(2)$ superscript in \eqn{fermionic energy} denotes
expansion to second order in perturbation theory. Recalling the
energy and spin values for the left, right, and massive modes
\eqn{boson mode weights} we find the total energies are given by
\begin{align}
\begin{split}
E_{L}&=\frac{1}{8\pi G\ell}\left(1-\frac{1}{\mu\ell}\right) \int
ed^{3}x\epsilon^{0\mu\nu}S_{\nu\mu0}^{(2)}
\left(\psi_{L}\right)\\
&\hphantom{=}+\frac{1}{32\pi G}\left(-1+\frac{1}{\mu\ell}\right)
\int ed^{3}x\bar{\nabla}^{0}h^{\mu\nu}_{L}\dot{h}_{L\mu\nu}\;,
\end{split}\\
E_{R}&=\frac{1}{32\pi G}\left(-1-\frac{1}{\mu\ell}\right)
\int ed^{3}x\bar{\nabla}^{0}h^{\mu\nu}_{R}\dot{h}_{R\mu\nu}\;,\\
\begin{split}
E_{M}&=\frac{1}{8\pi G\ell^{2}}\left(1-\mu\ell\right)
\left(2\mu\ell-1\right)\int ed^{3}x
\epsilon^{0\mu\nu}S_{\nu\mu0}^{(2)}\left(\psi_{M}\right)\\
&\hphantom{=}+\frac{1}{64\pi G} \left(\mu^{2}\ell^{2}-1\right)\int
ed^{3}x
\epsilon_{\beta}^{0\mu}h_{M}^{\beta\nu}\dot{h}_{M\mu\nu}\;.\end{split}
\end{align}
Taking into account that the  $\mathcal{N}=\left(1,0\right)$ and
$\mathcal{N}=\left(0,1\right)$ theories are related by parity, it is
easy to see that in the ${\cal N}=(0,1)$ theory the energies take
the form
\begin{align}
E_{L}&=\frac{1}{32\pi G}\left(-1-\frac{1}{\mu\ell}\right)
\int ed^{3}x\bar{\nabla}^{0}h^{\mu\nu}_{L}\dot{h}_{L\mu\nu}\;,\\
\begin{split}
E_{R}&=\frac{1}{8\pi G\ell}\left(1-\frac{1}{\mu\ell}\right)
\int ed^{3}x\epsilon^{0\mu\nu}S_{\nu\mu0}^{(2)}\left(\psi_{R}\right)\\
&\hphantom{=}+\frac{1}{32\pi G}\left(-1+\frac{1}{\mu\ell}\right)
\int ed^{3}x\bar{\nabla}^{0}h^{\mu\nu}_{R}\dot{h}_{R\mu\nu}\;,
\end{split}\\
\begin{split}
E_{M}&=\frac{1}{8\pi G\ell^{2}}\left(1-\mu\ell\right)
\left(2\mu\ell-1\right)\int
ed^{3}x\epsilon^{0\mu\nu}S_{\nu\mu0}^{(2)}
\left(\psi_{M}\right)\\
&\hphantom{=}+\frac{1}{64\pi G}\left(\mu^{2} \ell^{2}-1\right)\int
ed^{3}x\epsilon_{\beta}^{0\mu}h_{M}^{\beta\nu}\dot{h}_{M\mu\nu}\;.
\end{split}
\end{align}
At the linear level the corresponding expressions for the energies
of the ${\cal N}=(1,1)$ theory are given by
\begin{align}
E_{L} &=-\left(1 - \frac{1}{\mu\ell}\right)E_{B,L} + \left(1 -
\frac{1}{\mu\ell}\right)E_{F,L}\;,\\
E_{R} &=-\left(1 + \frac{1}{\mu\ell}\right)E_{B,R} - \left(1 -
\frac{1}{\mu\ell}\right)E_{F,R}\;,\\
E_{M} &=
\left(\mu^{2}\ell^{2}-1\right)E_{B,M}+\left(1-\mu\ell\right)\left(2\mu\ell-1\right)E_{F,M}\;,
\end{align}
where $E_{F,i}$ is as defined in \eqn{fermionic energy} with the
$\mu$ dependence factored out, and $E_{B,i}$ are the bosonic energies
as given in \cite{andy1}.
\ \\
\ \\
The evaluation of the fermionic energy
indicates that fermions do not contribute to quadratic order, though
we expect them to contribute to the next order in perturbation theory.
By the same reasoning as in the bosonic model \cite{andy1},
we conclude that for generic values of the coupling constants, the
energy is not positive semi-definite as one may naively expect for a
 supersymmetric theory. As elaborated in the introduction,
 it is known that when there are ghosts in
  the theory, their contribution to the energy will be negative, thereby spoiling the
   usual positivity arguments. Indeed, in concurrent work, Andrade
   and Marolf showed that TMG at general values of the coupling contains
    ghosts \cite{anma}.
   Similarly, as in the bosonic model, the energies of the
   ${\cal N}=(1,1)$ model
   become positive semi-definite at the chiral point $\mu \ell =1$ such that
\begin{equation}
E_{L} = E_{M} = 0,\quad E_{R} = 2E_{B}>0\;,
\end{equation}
in agreement with the disappearance of ghosts found in \cite{anma}.

\subsection{Fermionic Charges: Supercharge}
\label{supercharge subsection} 
Using the a perturbative expansion of
the gravitino equation of motion, we can calculate the supercharge to
order $\lambda^2$ so that
\begin{equation}
Q = \frac{1}{8\pi G}\int ed^{3}xJ_{0}\;.
\end{equation}
 Here $J_{0}$ is the temporal component of the supercurrent $J_{\mu}$,
 which is the gravitino
 field equation contracted with the appropriate Killing spinor. We begin by considering the $\mathcal{N}=\left(1,0\right)$ theory. As
 with the energy, we can calculate $Q$ to second order. 
 Schematically, the gravitino field equation takes the form
\begin{equation}
F_{\mu} = \lambda F^{(1)}_{\mu}(\psi) + \lambda^{2}
\left(F^{(1)}_{\mu}(\psi^{(2)}) + F^{(2)}_{\mu}(\psi)\right) +
\mathcal{O}(\lambda^{3})\;,
\end{equation}
with $F^{(1)\mu}$ given in \eqn{fermion field equation}.
$F^{(2)\mu}$ requires expanding the gravitino field equation to
order $\lambda^2$, and extracting the $\psi$ dependent terms. The
result is
\begin{align}
\begin{split}
F^{(2)\mu}(\psi)& = \epsilon^{\mu\nu\rho}\frac{1}{4}
\omega_{\nu}^{(1)mn}(h)\bar{\gamma}_{mn}\psi_{\rho} +
\frac{1}{2\ell}h^{\mu\nu}\psi_{\nu} -
\frac{i}{8\mu}\epsilon^{\sigma\nu\rho}\Big(R_{\sigma\nu}^{(1)\;ab}
(2\delta^{\mu}_{\rho}\gamma_{b}\psi_{a} + e^{\mu}_{b}
\bar{\gamma}_{\rho}\psi_{a})\\
&\hphantom{=} +
\frac{1}{\ell^2}(h^{\mu}_{\nu}\bar{\gamma}_{\rho}\psi_{\sigma} -
\delta^{\mu}_{\nu}h_{\rho\lambda}\bar{\gamma}^{\lambda}\psi_{\sigma}
+
\delta_{\nu}^{\mu}h_{\sigma\lambda}\bar{\gamma}_{\rho}\psi^{\lambda})\Big)
- \frac{i}{\mu}\left(\frac{1}{2\ell}R^{(2)\mu} -
\frac{1}{4\ell^2}h^{\mu\nu}\psi_{\nu}\right)\;,
\end{split}
\end{align}
where
\begin{equation}
R^{(2)\mu}(\psi) =
\frac{1}{4}\epsilon^{\mu\nu\rho}\omega_{\nu}^{(1)nm}(h)\bar{\gamma}_{mn}\psi_{\rho}\;,
\end{equation}
and
\begin{equation}
R_{\mu\nu ab}^{(1)} =
(R_{\mu\nu\alpha\beta}e^{\alpha}_{a}e^{\beta}_{b})^{(1)}\;.
\end{equation}
After plugging in
 the first-order modes, we find  the $\mathcal{N}=\left(1,0\right)$ supercharges (to
order $\lambda^{2}$)
\begin{equation}
Q_{L}= \left(1 - \frac{1}{\mu\ell}\right)Q_{\omega,L}+\left(1 -\frac{1}{\mu\ell}\right)Q_{h,L}
\end{equation}
and
\begin{equation}
Q_{M}= \left(1 - \frac{1}{\mu\ell}\right)Q_{\omega,M}+\left(1 -\frac{1}{\mu\ell}\right)Q_{h,M}\;.
\end{equation}
At the chiral point, $\mu\ell = 1$, we have
\begin{equation}
Q_{L} = 0\quad \hbox{and}\quad Q_{M}= 0\;.
\end{equation}
In the above formulas, we defined
\begin{align}
Q_{\omega}\left(\xi,\psi,h\right)&=\frac{1}{32\pi G}\int e d^3x
\bar{\xi}\left(\epsilon^{0\nu\rho}\omega_{\nu}^{(1)mn}(h)
\bar{\gamma}_{mn}\psi_{\rho}\right)\;,\\
Q_{h}\left(\xi,\psi,h\right)&=\frac{1}{32\pi G\ell}\int e d^3x
  \bar{\xi}\epsilon^{0\nu\rho}h_{\nu\lambda}
  \bar{\gamma}^{\lambda}\psi_{\rho}\;,\\
Q_{\omega,i}&=Q_{\omega}\left(\xi_{i},\psi_{i},h_{i}\right),\\
Q_{h,i}&=Q_{h}\left(\xi_{i},\psi_{i},h_{i}\right)\;,
\end{align}
where the subindex labels individual modes.
\ \\
\ \\
Applying a parity transformation leads to the supercharges of the
${\cal N}=(0,1)$ theory:
\begin{equation}
Q_{R}=\left(1 - \frac{1}{\mu\ell}\right)Q_{\omega,R}-\left(1 -\frac{1}{\mu\ell}\right)Q_{h,R}\;,
\end{equation}
and
\begin{equation}
Q_{M}=\left(1 - \frac{1}{\mu\ell}\right)Q_{\omega,M}-\left(1 -\frac{1}{\mu\ell}\right)Q_{h,M}\;.
\end{equation}
At the chiral point, $\mu\ell = 1$, we have
\begin{equation}
Q_{M}= 0\quad\hbox{and}\quad Q_{R} =0\;.
\end{equation}
Similarly, at the linear level ${\cal N}=(1,1)$, charges are found to
be
\begin{align}
Q_{L}&=\left(1 -\frac{1}{\mu\ell}\right)Q_{\omega,L}+\left(1 -\frac{1}{\mu\ell}\right)Q_{h,L}\;,\\
Q_{R}&=\left(1 +\frac{1}{\mu\ell}\right)Q_{\omega,R}-\left(1 +\frac{1}{\mu\ell}\right)Q_{h,R}\;,\\
Q_{M}&=\left(1 -\frac{1}{\mu\ell}\right)Q_{\omega,M}+\left(1 -\frac{1}{\mu\ell}\right)Q_{h,M}\;.
\end{align}
At the chiral point only the right moving supercharge is
non-vanishing
\begin{align}
Q_{L}= 0\;,\quad Q_{R}=2Q_{\omega,R}-2Q_{h,R}\;,\quad\hbox{and}\quad Q_{M}=0\;.
\end{align}
Therefore, the fermionic charges share the bosonic charges' chiral
behavior.

\section{Conclusion}
\label{conclusion}

In this paper we have constructed Chiral Supergravity, the ${\cal
N}=1$ supersymmetric extension of Chiral Gravity. The theory has
been studied in a perturbative regime around the AdS$_{3}$
background. The wave functions have been constructed, and the conserved charges were computed to second order. These charges picked out a
distinguished point in parameter space, $\mu\ell=1$, at which the
theory acquires a chiral nature in a fashion similar to its bosonic
counterpart \cite{andy1}.
\ \\
\ \\
Although the positivity of energy could not be proven, at the chiral
  point, it was shown that AdS$_{3}$ is stable against
  perturbations. Thus, Chiral Supergravity is a consistent theory at the
classical level.
\ \\
\ \\
   Several important questions remain: Given
   the existence of a classical supersymmetric extension of Chiral Gravity, it would be interesting to examine if the
quantum theory is consistent. Also, given the calculation of
the partition function \cite{Maloney}, as well as the recent claim
about renormalizability of TMG \cite{ioda}, it seems likely that
Chiral Supergravity is a consistent theory of gravity even at the
quantum level. Some more work on the supersymmetric theory needs to
be done to confirm this. 
\ \\
\ \\
If Chiral Supergravity is consistent at the quantum level, it would be of value to see whether or not it is derivable from String Theory.  Some work in this direction was done by Gupta and Sen\cite{guptasen}. They found a consistent truncation of higher dimensional Supergravity with
matter fields to pure three-dimensional, cosmological Supergravity with a gravitational Chern-Simons term.
Their truncation involves a scale hierachy, and the Chern-Simons coupling is assumed to be smaller than the (anti) de Sitter radius. In Chiral Supergravity, unfortunately, these paramteres are equal. Given that Chiral Supergravity is three-dimensional, an alternative
way to find a relation to String Theory might be to map Chiral
Supergravity to a string sigma model.
\ \\
\ \\
   It would also be interesting to compute the partition function for
   Chiral Supergravity. Due to the recent work of Maloney, Song and Strominger \cite{Maloney}, one may
   anticipate this partition function to correspond to the chiral part of the
   ${\cal N}=1$ partition function calculated in \cite{Maloney2}.
\ \\
\ \\
   In light of the recent work of Gaberdiel et. al. \cite{gagu}, where difficulties in
   constructing ${\cal N}=2$ extremal conformal field theories with large
central
   charge were reported, it would be interesting to construct an
   ${\cal N}=2$ version of TMG. Kaura and Sahoo had one attempt\cite{sahoo}, but clearly more work needs to be done.
\ \\
\ \\
Though we focused on the supersymmetric
version of Chiral Gravity, there are indications for the existence
of a supersymmetric version of Log Gravity. As in the
bosonic case\cite{grjo1,henneaux2}, there is an additional solution to the gravitino equation of motion at the chiral point
\begin{equation*}
\psi_{\mu}=y(\tau,\rho)\psi_{\mu},
\end{equation*} with $y=-i\tau-\ln\cosh\rho$.
 The fermionic boundary terms of the action are needed to verify that this mode obeys the variational
 principle and better specify the boundary conditions. This would allow for a detailed study of Log Supergravity and a possible
  dual logarithmic superconformal field theory. 
  Fermionic boundary conditions have been discussed in \cite{compere} and references therein. We hope to address some of these questions in the future.
\newpage
\acknowledgments
It is a pleasure to thank T.~Andrade, G.~Comp\`{e}re, D.~Grumiller,
M.~Henneaux, C.~Keller, W.~Li, D.~Marolf, A.~Maloney, M.~Rocek,
W.~Song and A.~Strominger for
  useful discussions. We thank E.~Sezgin for his collaboration at
  the early stages of this project. Many thanks to the Erwin
  Schr\"odinger Institute (Vienna)
  and to the organizers of the ``Workshop on 3D Gravity'' (April 09)
  for a very exciting meeting where this work was presented.
  We especially would like to thank the KITP Santa Barbara for uts kind
  hospitality during the program
  ``Fundamental Aspects of Superstring Theory'' where this project was
carried out.
  This work was supported by NSF under grant PHY-0505757 and the
  University of Texas A\&M. M.~B. would like to thank KITP Santa
  Barbara for partial financial support under National Science Foundation 
  under Grant No. PHY05-51164.

\appendix
\section{Appendix: Notation and Conventions}
\label{notation and conventions appendix}

Throughout the course of this paper we have worked with an index
structure such that flat coordinates are labeled by Latin indices,
$a,\;b,\ldots=0,\;1,\;2$ and curved coordinates are labeled by Greek
indices, $\mu,\;\nu,\ldots=0,\;1,\;2$. Moreover, the manifold
parameterized by the flat coordinates is endowed with a Minkowski
metric $\eta$ of signature $\left(-,+,+\right)$. The
 background curved space is taken to
be $AdS_{3}$ in global coordinates:
\begin{equation*}
ds^{2}=\bar{g}_{\mu\nu}dx^{\mu}dx^{\nu}=\ell^{2}\left(-\cosh^{2}
\rho d\tau^{2}+\sinh^{2}\rho d\phi^{2}+d\rho^{2}\right)\;,
\end{equation*}
and the Lorentzian theory is considered so that only $\phi$ is
cyclic. In such a geometry, one can speak of a conformal boundary understood to lie at $\rho=\infty$. Moreover, we chose a
space-time orientation such that
\begin{equation*}
\epsilon_{012}=-\epsilon^{012}=+1\;,\quad
\epsilon^{\mu\alpha\beta}\epsilon_{\mu\lambda\sigma}
=-\delta^{\alpha\beta}_{\lambda\sigma}\;,\quad\hbox{and}\quad
\epsilon^{\mu\alpha\beta}
\epsilon_{\mu\alpha\sigma}=-2\delta^{\beta}_{\sigma}\;.
\end{equation*}
The theories under consideration all involve spinors which are taken
to lie in the Majorana representation. Moreover all fermions are
implicitly Grassmann-valued spinors. Taking $\chi$ and $\phi$ to be
Grassmann-valued Majorana spinors, we have the following useful
identities:
\begin{equation*}
\bar{\chi}\phi=\bar{\phi}\chi\;,\quad
\bar{\chi}\gamma_{\mu}\phi=-\bar{\phi}\gamma_{\mu}\chi\;,\quad
\bar{\chi}\gamma_{\mu}\gamma_{\nu}\phi=\bar{\phi}\gamma_{\nu}\gamma_{\mu}\chi\;,
\end{equation*}
where $\gamma_{\mu}=e^{a}_{\mu}\gamma_{a}$ are three-dimensional
gamma matrices and bar denotes the Dirac adjoint
$\bar{\chi}=\chi^{\dagger}\gamma_{0}$, with $\chi^{\dagger}=
\left(\chi^{*}\right)^{T}$.
\ \\
\ \\
The curvatures and connections are given by
\begin{align*}
\omega_{\mu}^{ab}=e^{a}_{\nu}\partial_{\mu}e^{\nu
b}+e^{a}_{\nu}e^{\sigma b} \Gamma^{\nu}_{\mu\sigma}\;,\quad
\Gamma^{\lambda}_{(\mu\nu)}
=\frac{1}{2}g^{\lambda\rho}\left(\partial_{\mu}g_{\rho\nu}+\partial_{\nu}
g_{\mu\rho}-g_{\rho}g_{\mu\nu}\right)\;,\\
R=e_{a}^{\mu}e^{\nu}_{b}R^{ab}_{\mu\nu}\left(\omega\right),\quad\hbox{and}\quad
R_{\mu\nu}{}^{ab}
=\partial_{\mu}\omega_{\nu}^{ab}-\partial_{\nu}\omega_{\mu}^{ab}+
\omega_{\mu}^{ac}\omega_{\nu c}^{b}-\omega_{\nu}^{ac}\omega_{\mu
c}^{b}\;.
\end{align*}
Covariant derivatives acting on spinors $\epsilon$, vector-spinors
$\psi_{\mu}$, and tensors are
\begin{align*}
D_{\mu}\epsilon=\partial_{\mu}\epsilon+\frac{1}{4}\omega_{\mu}^{ab}
\gamma_{ab}\epsilon\;,\quad
D_{\alpha}\psi_{\beta}
=\partial_{\alpha}\psi_{\beta}+\frac{1}{4}\omega^{ab}_{\alpha}
\gamma_{ab}\psi_{\beta}-\Gamma^{\lambda}_{\alpha\beta}\psi_{\lambda}\;,\quad\hbox{and}\\
\nabla_{\mu}X_{\nu\rho}=\partial_{\mu}X_{\nu\rho}-
\Gamma_{\mu\nu}^{\sigma}X_{\sigma\rho}-\Gamma_{\mu\rho}^{\sigma}X_{\nu\sigma}\;.
\end{align*}
Upon a perturbative expansion of the metric \eqn{metric
perturbation}
 and the Rarita-Schwinger field
\eqn{rarita schwinger perturbation}, we find the tensors of the
 $\mathcal{N}=\left(1,0\right)$ theory:
\begin{align*}
R_{\mu\nu\alpha\beta}^{(1)}&= \frac{1}{2\ell^{2}}
(h_{\mu\alpha}\bar{g}_{\nu\beta}) - h_{\nu\beta}\bar{g}_{\mu\alpha})\;,\\
\bar{R}_{\mu\nu}{}^{ab}&=\Lambda\left(e^{a}_{\mu}e^{b}_{\nu}-e^{b}_{\mu}e^{a}_{\nu}\right)\;,\\
\bar{R}_{\mu\nu}&=2\Lambda\bar{g}_{\mu\nu}\;,\\
\bar{R}&=6\Lambda,\\
R^{(1)}&=-\bar{\nabla}^{2} h +
\bar{\nabla}_{\mu}\bar{\nabla}_{\nu}h^{\mu\nu} - 2\Lambda h\;,\\
\mathcal{G}^{(1)}_{\mu\nu}\left(h\right)&=R^{(1)}_{\mu\nu}
\left(h\right)-\frac{1}{2}\bar{g}_{\mu\nu}R^{(1)}\left(h\right)-
\Lambda h_{\mu\nu}\;,\\
\mathcal{G}^{(2)}_{\mu\nu}\left(h,\psi\right)&=\mathcal{G}^{(2)}_{\mu\nu}
\left(h\right)+\bar{\nabla}_{\rho}\kappa^{(2)}_{\mu\nu}{}^{\rho}\left(\psi\right)\;,\\
\mathcal{G}^{(2)}_{\mu\nu}\left(h\right)&=R^{(2)}_{\mu\nu}
\left(h\right)-\frac{1}{2}\bar{g}_{\mu\nu}R^{(2)}\left(h\right)\;,\\
R_{\mu\nu}^{(2)}\left(h\right)&=\frac{1}{4}\bar{\nabla}_{\mu}h_{\alpha\beta}
\bar{\nabla}_{\nu}h^{\alpha\beta}
+\bar{\nabla}_{[\beta}h_{\alpha]\mu}\bar{\nabla}^{\beta}h^{\alpha}{}_{\nu}
+h^{\alpha\beta}\left(\bar{\nabla}_{\alpha}\left(\bar{\nabla}_{[\beta}
h_{\mu]\nu}\right)+\bar{\nabla}_{\nu}\left(\bar{\nabla}_{[\mu}h_{\alpha]\beta}\right)\right)\;,\\
R^{(2)}\left(h\right)&=\bar{g}^{\mu\nu}R^{(2)}_{\mu\nu}
\left(h\right)-\frac{3}{\ell^{2}}h^{\mu\nu}h_{\mu\nu}\;,\\
C^{(2)}_{\mu\nu}\left(h\right)&=\frac{1}{2}\left[\epsilon_{\mu}{}^{\alpha\beta}
\bar{\nabla}_{\alpha}\mathcal{G}^{(2)}_{\beta\nu}\left(h\right)
+h_{\mu\lambda}\epsilon^{\lambda\alpha\beta}\mathcal{G}_{\beta\nu}^{(2)}-
\frac{h}{2}\epsilon_{\mu}{}^{\alpha\beta}\bar{\nabla}
\mathcal{G}^{(1)}_{\beta\nu}\left(h\right)-\epsilon_{\mu}{}^{\alpha\beta}\Gamma^{(1)
\lambda}_{\nu\alpha}\mathcal{G}_{\beta\lambda}^{(1)}
\left(h\right)+\left(\mu\leftrightarrow\nu\right)\right]\;,\\
\Gamma^{(1)\lambda}_{\mu\nu}\left(h\right)&=\frac{1}{2}\left(\bar{\nabla}_{\mu}
h^{\lambda}_{\nu}+\bar{\nabla}_{\nu}h^{\lambda}_{\mu}
-\bar{\nabla}^{\lambda}h_{\mu\nu}\right)\;,\\
\Gamma^{(2)\lambda}_{\mu\nu}\left(h\right)&=-\frac{1}{2}h^{\lambda\rho}
\left(\bar{\nabla}_{\mu}h_{\rho\nu}+\bar{\nabla}_{\nu}h_{\rho\mu}
-\bar{\nabla}_{\rho}h_{\mu\nu}\right)\;,\\
C^{(2)}_{\mu\nu}\left(h,\psi\right)&=C_{\mu\nu}^{(2)}\left(h\right)+
\epsilon_{\mu}{}^{\alpha\beta}\bar{\nabla}_{\alpha}\bar{\nabla}_{\rho}
\kappa^{(2)}_{\beta\nu}{}^{\rho}\left(\psi\right)-\frac{i}{8}\Lambda
\epsilon_{\mu}{}^{\alpha\beta}\bar{\psi}_{\alpha}\bar{\gamma}_{\nu}\psi_{\beta}\;,\\
F^{(2)}_{\mu\nu}\left(\psi\right)&=\frac{i}{4\ell\bar{e}}\epsilon^{\sigma
\lambda\rho}\bar{g}_{\lambda\mu}\bar{\psi}_{\sigma}\bar{\gamma}_{\nu}
\psi_{\rho}-\frac{i}{8\mu\ell^{2}\bar{e}}\epsilon^{\rho\sigma\tau}
\bar{g}_{\mu\tau}\bar{\psi}_{\rho}\bar{\gamma}_{\nu}\psi_{\sigma}\;,\\
&\hphantom{=}-\frac{i}{2\mu\ell^{2}\bar{e}}\left(1-E\right)^{2}
\left(\bar{g}_{\mu\nu}\epsilon^{\rho\sigma\tau}\bar{\psi}_{\sigma}
\bar{\gamma}_{\tau}\psi_{\rho}-2\bar{g}_{\mu\sigma}\epsilon^{\sigma\rho\tau}
\bar{\psi}_{\rho}\bar{\gamma}_{\tau}\psi_{\nu}\right)\;,\\
\kappa^{(2)}_{\mu\nu}{}^{\rho}\left(\psi\right)&=\frac{i}{4}
\left(\bar{\psi}_{\mu}\bar{\gamma}_{\nu}\psi^{\rho}-\bar{\psi}_{\mu}
\bar{\gamma}^{\rho}\psi_{\nu}+\bar{\psi}_{\nu}\bar{\gamma}_{\mu}\psi^{\rho}\right)\;.\\
\end{align*}
The vielbein decomposes as
\begin{equation*}
e^{a}_{\mu}=\bar{e}^{a}_{\mu} + \lambda e^{(1)a}_{\mu} +
\mathcal{O}\left(\lambda^{2}\right)\;.
\end{equation*}
Employing the relationship between the metric and the vielbein along
with our linearized metric solution we find
\begin{align*}
e^{(1)0}_{\tau} &= -\frac{h_{00}}{2\cosh\rho}\;, & e^{(1)0}_{\phi} &=
-\frac{h_{01}}{2\cosh\rho}\;,  & e^{(1)0}_{\rho} &=
-\frac{h_{02}}{2\cosh\rho}\;,\\
e^{(1)1}_{\tau} &= \frac{h_{01}}{2\sinh\rho}\;, & e^{(1)1}_{\phi} &=
\frac{h_{11}}{2\sinh\rho}\;,
 & e^{(1)1}_{\rho} &= \frac{h_{12}}{2\sinh\rho}\;, \\
e^{(1)2}_{\tau} &= \frac{1}{2}h_{02}\;, & e^{(1)2}_{\phi} &=
\frac{1}{2}h_{12 }\;,\quad\hbox{and} & e^{(1)2}_{\rho}&= \frac{1}{2}h_{22}\;.
\end{align*}
The linearized spin connection in terms of the vielbeins is
\begin{equation*}
\begin{split}
\omega_{\mu ab}^{(1)}&= -\frac{1}{2}e^{(1)\nu}_{a}
(\partial_{\mu}\bar{e}_{b \nu} -
 \partial_{\nu}\bar{e}_{\mu b}) +
\frac{1}{2}\bar{e}^{\nu}_{a}(\partial_{\mu}e^{(1)}_{b\nu} -
  \partial_{\nu}e^{(1)}_{\mu b}) - \frac{1}{2}\bar{e}^{\alpha}_{a}\bar{e}^{\beta}_{b}
  \bar{e}_{\mu}^{c}(\partial_{\alpha}e^{(1)}_{\beta c})\\
&\hphantom{=}- \frac{1}{2}\bar{e}^{\alpha}_{a}\bar{e}^{\beta}_{b}e_{\mu}^{(1)c}
(\partial_{\alpha}\bar{e}_{\beta c}) + \frac{1}{2}e_{a}^{(1)\alpha}
\bar{e}^{\beta}_{b}\bar{e}_{\mu}^{c}(\partial_{\alpha}\bar{e}_{\beta c}) + \frac{1}{2}
\bar{e}^{\alpha}_{a}e^{(1)\beta}_{b}\bar{e}^{c}_{\mu}(\partial_{\alpha}\bar{e}_{\beta
c}) - (a\leftrightarrow b)\;.
\end{split}
\end{equation*}
It is useful to note
\begin{equation*}
\left[\bar{\nabla}_{\sigma},\bar{\nabla}_{\mu}\right]
h^{\sigma}_{\mu}=\bar{R}^{\sigma}{}_{\lambda\sigma\mu}h^{\lambda}_{\nu}
-\bar{R}^{\lambda}{}_{\nu\sigma\mu}h^{\sigma}_{\lambda}=3\Lambda
h_{\mu\nu}
-\Lambda h\bar{g}_{\mu\nu}
\end{equation*}
and
\begin{equation*}
\left[D_{\mu},D_{\nu}\right]\psi_{a}=R_{\mu\nu
ab}\psi^{b}+\frac{1}{4}R_{\mu\nu bc} \gamma^{bc}\psi_{a}\;.
\end{equation*}
At the linear level, the fermionic field strength of the
$\mathcal{N}=\left(1,0\right)$ theory satisfies
\begin{equation*}
f^{(1)}_{\mu}=\frac{1}{\ell}\left(E-1\right)\psi_{\mu}\;.
\end{equation*}

\section{Clifford Algebras, Spinor Representations, and Discrete Symmetries}
\label{Clifford algebra appendix} In TMG, a parity transformation
effectively takes $\mu\rightarrow -\mu$. When
 dealing with fermions in
TMSG, we must also consider the operator $\mathcal{P}$ acting on
fermions.

\subsection{Clifford Algebra and Spinor Representation}
\label{clifford algebra subappendix}

The structure of the gravitino fields descends from defining spinors
on the global group of isometries, $SO(2,2)$. This is the isometry
group of $\mathbb{R}^{2+2}$ of which $AdS_{3}$ is a hypersurface.
The Clifford algebra of gamma matrices is
\begin{equation}
\label{capital gamma}
\left\{\Gamma_{A},\Gamma_{B}\right\} = 2\eta_{AB}\;.
\end{equation}
Here the metric is $\eta = \mathrm{diag}(-1,-1,1,1)$ and $A =
-,0,1,2$. The ``$-$'' index represents the additional direction in
$\mathbb{R}^{2+2}$, so that the gamma matrices explicitly read:
\begin{equation*}
\Gamma_{A} = \begin{pmatrix}0 & \gamma_{A} \cr
\hat{\gamma}_{A} & 0\end{pmatrix}\;,
\end{equation*}
with
\begin{equation*}
\gamma_{0} = \begin{pmatrix}0&1\cr -1&0\end{pmatrix}\;,
\quad \gamma_{1} = \begin{pmatrix}0&1\cr 1&0\end{pmatrix}\;,
\quad\hbox{and}\quad \gamma_{2} =
\begin{pmatrix}1&0\cr 0&-1\end{pmatrix}\;,
\end{equation*}
$\gamma_{-} = \hat{\gamma}_{-} = \mathbb{I}$, and $\hat{\gamma}_{m}
= -\gamma_{m}$, for $m=0,1,2$. The hypersurface in
$\mathbb{R}^{2+2}$ is a curved space. We employ the dreibein
$e_{\mu}^{m}$ to define $\gamma_{\mu}$ on this curved space:
\begin{equation*}
\gamma_{\mu} = e_{\mu}^{m}\gamma_{m}\;.
\end{equation*}
Three-dimensional gamma matrices satisfy
\begin{equation*}
\{\gamma_{\mu}\gamma_{\nu}\} = 2g_{\mu\nu}\;,
\end{equation*}
and in addition
\begin{equation*}
[\gamma_{\mu},\gamma_{\nu}] = \epsilon_{\mu\nu\rho}\gamma^{\rho}\;,
\quad [\hat{\gamma}_{\mu},\hat{\gamma}_{\nu}] =
-\epsilon_{\mu\nu\rho}\hat{\gamma}^{\rho}\;,
\end{equation*}
where again, with curved indices $\epsilon_{012} =
\ell\sinh\rho\cosh\rho$. We have used the notation
\begin{equation*}
\gamma^{\mu\nu} = \frac{1}{2!}(\gamma^{\mu}\gamma^{\nu} -
\gamma^{\nu}\gamma^{\mu})\;, \quad \hbox{and} \quad
\gamma^{\mu\nu\rho} = \gamma^{[\mu}\gamma^{\nu}\gamma^{\rho]} =
\epsilon^{\mu\nu\rho}.
\end{equation*}
Four component ``Dirac'' spinors are decomposed into two component
Weyl spinors. These Weyl spinors have the correct dimensionality for
spinor fields in three dimensions. Fermions are taken to be Majorana
spinors and have anticommuting components.

\subsection{Discrete Symmetries}
\label{discrete symmetries subappendix} The Dirac notation is
convenient for discussing discrete symmetries. We use the block
diagonal form from equation \eqn{capital gamma}. If we define a Dirac spinor on $AdS_{3}$
by
\begin{equation*}
\Psi = \begin{pmatrix}\psi^{R} \cr \psi^{L}\end{pmatrix}\;,
\end{equation*}
then the parity operator acts on this spinor as
\begin{equation*}
\mathcal{P}\Psi = i\Gamma_{1}\Psi\;.
\end{equation*}
In Weyl language, we have
\begin{equation*}
\mathcal{P}\psi^{R}_{\mu} = -i\gamma_{1}\psi^{R}_{\mu} = \psi^{L}_{\mu}
\end{equation*}
and
\begin{equation*}
\mathcal{P}\psi^{L}_{\mu} = i\gamma_{1}\psi^{L}_{\mu} =
\psi^{R}_{\mu}\;.
\end{equation*}
It is also useful to define the charge conjugation operator
$\mathcal{C}$ using Dirac notation. In this case:
\begin{equation*}
\mathcal{C} = \begin{pmatrix}\hat{\gamma}_{0}& 0 \cr 0 &
\gamma_{0}\end{pmatrix}\;.
\end{equation*}
In Weyl language, we have
\begin{align*}
\mathcal{C}\psi^{R}_{\mu} = -\gamma_{0}\psi^{R}_{\mu}\;,\\
\mathcal{C}\psi^{L}_{\mu} = \gamma_{0}\psi^{L}_{\mu}\;.
\end{align*}
The Majorana condition is then
\begin{equation*}
\bar{\Psi} = \Psi^{T}\mathcal{C}\;.
\end{equation*}

\nocite{*}
\bibliography{CTMSG-JHEP-v3}{}
\bibliographystyle{JHEP}
\end{document}